\documentclass[11pt, a4paper]{article}
\pdfoutput=1

\usepackage{jcappub}
\usepackage{amsmath}
\usepackage{textcomp}

\title{Cosmic-muon flux and annual modulation in Borexino at 3800 m water-equivalent depth}

\collaboration{Borexino collaboration}

\author[a]{G.~Bellini,}
\author[f]{J.~Benziger,}
\author[q]{D.~Bick,}
\author[b]{G.~Bonfini,}
\author[c]{D.~Bravo,}
\author[a]{M.~Buizza Avanzini,}
\author[a]{B.~Caccianiga,}
\author[p]{L.~Cadonati,}
\author[d]{F.~Calaprice,}
\author[g]{C.~Carraro,}
\author[b]{P.~Cavalcante,}
\author[d]{A.~Chavarria,}
\author[r]{A.~Chepurnov,}
\author[a]{D.~D{\textquoteright}Angelo,} 
\author[g,s]{S.~Davini,}
\author[i]{A.~Derbin,}
\author[j]{A.~Etenko,}
\author[h]{F.~von~Feilitzsch,}
\author[k]{K.~Fomenko,}
\author[e]{D.~Franco,}
\author[d]{C.~Galbiati,}
\author[b]{S.~Gazzana,}
\author[b]{C.~Ghiano,}
\author[a]{M.~Giammarchi,}
\author[h]{M.~Goeger-Neff,}
\author[d]{A.~Goretti,}
\author[d]{L.~Grandi,}
\author[g]{E.~Guardincerri,}
\author[q]{C.~Hagner,}
\author[c]{S.~Hardy,}
\author[b]{Aldo~Ianni,}
\author[d]{Andrea~Ianni,}
\author[k]{D.~Korablev,}
\author[s]{G.~Korga,}
\author[b]{Y.~Koshio,}
\author[e]{D.~Kryn,}
\author[b]{M.~Laubenstein,}
\author[h]{T.~Lewke,}
\author[j]{E.~Litvinovich,}
\author[d]{B.~Loer,}
\author[b]{F.~Lombardi,}
\author[a]{P.~Lombardi,}
\author[a]{L.~Ludhova,}
\author[j]{I.~Machulin,}
\author[c]{S.~Manecki,}
\author[l]{W.~Maneschg,}
\author[g]{G.~Manuzio,}
\author[h]{Q.~Meindl,}
\author[a]{E.~Meroni,}
\author[a]{L.~Miramonti,}
\author[b,n]{M.~Misiaszek,} 
\author[b,d]{D.~Montanari,}
\author[d]{P.~Mosteiro,}
\author[i]{V.~Muratova,}
\author[h]{L.~Oberauer,}
\author[e]{M.~Obolensky,}
\author[o]{F.~Ortica,}
\author[p]{K.~Otis,}
\author[g]{M.~Pallavicini,} 
\author[b]{L.~Papp,}
\author[g]{L.~Perasso,}
\author[g]{S.~Perasso,}
\author[p]{A.~Pocar,}
\author[c]{R.S.~Raghavan,}
\author[a]{G.~Ranucci,}
\author[b]{A.~Razeto,}
\author[a]{A.~Re,}
\author[o]{A.~Romani,}
\author[j]{A.~Sabelnikov,}
\author[d]{R.~Saldanha,} 
\author[b]{C.~Salvo,}
\author[h]{S.~Sch\"onert,}
\author[l]{H.~Simgen,}
\author[j]{M.~Skorokhvatov,}
\author[k]{O.~Smirnov,}
\author[k]{A.~Sotnikov,} 
\author[j]{S.~Sukhotin,}
\author[b]{Y.~Suvorov,}
\author[b]{R.~Tartaglia,}
\author[g]{G.~Testera,}
\author[e]{D.~Vignaud,}
\author[c]{R.B.~Vogelaar,} 
\author[h]{J.~Winter,}
\author[n]{M.~Wojcik,}
\author[d]{A.~Wright,}
\author[q]{M.~Wurm,}
\author[d]{J.~Xu,}
\author[k]{O.~Zaimidoroga,}
\author[g]{S.~Zavatarelli,}
\author[n]{G.~Zuzel}

\affiliation[a]{Dipartimento di Fisica, Universit\`a di Milano and INFN Milano, via Celoria 16, I-20133 Milano, Italy}

\affiliation[b]{Laboratori Nazionali del Gran Sasso, SS 17bis Km 18+910, I-67010 Assergi (AQ), Italy}

\affiliation[c]{Physics Department, Robeson Hall, Virginia Polytechnic Institute and State University, Blacksburg, VA 24061-0435, USA} 

\affiliation[d]{Department of Physics, Princeton University, Jadwin Hall, Washington Road, Princeton, NJ 08544-0708, USA}

\affiliation[e]{Astroparticule et Cosmologie APC, 10 rue Alice Domon et L\'eonie Duquet, 75205 Paris cedex 13, France}

\affiliation[f]{Department of Chemical Engineering, Princeton University, Engineering Quadrangle, Princeton, NJ 08544-5263, USA}

\affiliation[g]{Dipartimento di Fisica, Universit\`a di Genova and INFN Genova, via Dodecaneso 33, I-16146 Genova, Italy}

\affiliation[h]{Technische Universit\"at M\"unchen, James Franck Strasse E15, D-85747 Garching, Germany}

\affiliation[i]{St. Petersburg Nuclear Physics Institute, Gatchina, Russia}

\affiliation[j]{NRC Kurchatov Institute, Kurchatov Sq. 1, 123182 Moscow, Russia}

\affiliation[k]{JINR, Joliot Curie str. 6, 141980 Dubna, Russia}

\affiliation[l]{Max-Planck-Institut f\"ur Kernphysik, Postfach 103 980, D-69029 Heidelberg, Germany}

\affiliation[n]{Institute of Physics, Jagiellonian University, ul. Reymonta 4, PL-30059 Krakow, Poland}

\affiliation[o]{Dipartimento di Chimica, Universit\`a di Perugia and INFN Perugia, via Elce di Sotto 8, I-06123 Perugia, Italy}

\affiliation[p]{Physics Department, University of Massachusetts, Amherst, MA 01003, USA}

\affiliation[q]{Institut f\"ur Experimentalphysik, Universit\"at Hamburg, 22761 Hamburg, Germany}

\affiliation[r]{Institute of Nuclear Physics, Lomonosov Moscow State University, 119899, Moscow, Russia}

\affiliation[s]{Department of Physics, University of Houston, 4800 Calhoun Rd, Houston,  TX  77204, USA}

\abstract{
We have measured the muon flux at the underground Gran Sasso National Laboratory (3800\,m\,w.e.) 
to be $(3.41\pm0.01)\cdot10^{-4}$m$^{-2}$s$^{-1}$ using four years of Borexino data. 
A modulation of this signal is  observed with a period of ($366\pm3$) days and a relative amplitude of ($1.29\pm0.07$)\%. 
The measured phase is ($179\pm6$) days, corresponding to a maximum on the 28$^{\rm th}$ of June.  
Using the most complete atmospheric data models available, 
muon rate fluctuations are shown to be positively correlated with atmospheric temperature, 
with an effective coefficient $\alpha_T = 0.93 \pm 0.04$. 
This result represents the most precise study of the muon flux modulation for this site and is in good agreement with expectations.
}


\arxivnumber{1202.6403}

\notoc

\begin{document}
\maketitle

\section{Introduction} 
\label{sec:intro}

Muons observed in underground sites arise mostly from the decay of pions
and kaons produced by the interaction of primary cosmic rays with the nuclei
of the upper atmosphere \cite{gai90}. 
Since muons lose energy as they penetrate the Earth, 
low-energy muons are stopped and only the most energetic muons can be
observed in underground detectors, with an energy threshold, $E_{\textrm{thr}}$, depending
on the depth. 
The flux of cosmic muons detected deep underground shows
time variations which are, at first approximation, seasonal. These variations can
be related to the air density fluctuations, which affect the fraction of
mesons decaying to muons energetic enough to reach the underground detector.
This effect has been known and studied for many decades \cite{bar52}. 
Underground experiments have observed this phenomenon at Gran Sasso (MACRO \cite{macro97}, LVD \cite{sel09}) and in other underground sites (\cite{icecube}, \cite{minos10} and refs. therein). 

Borexino is an organic liquid scintillator detector \cite{bx08det} located in the underground Gran Sasso National Laboratory (LNGS) in central Italy under a limestone coverage of $\sim$1300\,m ($\sim$3800\,m\,w.e). 
It is devoted to the spectroscopy of low-energy solar neutrinos via elastic scattering on electrons. 
Data taking started in May 2007 and led to measurements of solar neutrinos ($^7$Be \cite{bx11be7, bx12dn}, pep \cite{bx11pep}, $^8$B \cite{bx10b8}, and a limit on CNO \cite{bx11pep}), as well as antineutrinos from the Earth (geo-neutrinos) \cite{bx10geo}.
Borexino is also a powerful tool for both the study of cosmic muons that penetrate the Gran Sasso rock coverage and the neutrons and radioactive isotopes that they produce, which are relevant backgrounds for dark matter and neutrino experiments. 

Borexino can select muons passing through a spherical volume with a cross section of 146\,m$^2$.
Such a geometry makes the acceptance independent of the muon angle of incidence, 
allowing us to measure the muon flux and its time dependence with reduced systematics. 
Furthermore, as air temperature data can be obtained from specialized atmosphere modeling centers for weather forecast \cite{ecmwf}, the correlation with the muon flux can be investigated and the effective temperature coefficient can be determined.

In this article we first introduce the Borexino detector (section~\ref{sec:bx}) and report on the measured muon flux (section~\ref{sec:flux}) and on its seasonal modulation (section~\ref{sec:modulation}). 
We then briefly describe how the muon flux is expected to be related to the atmospheric effective temperature (section~\ref{sec:model}), 
before reporting the temperature fluctuations at LNGS (section~\ref{sec:temp}). 
Both muon flux and temperature modulations are also analyzed with Lomb-Scargle frequency analysis (section~\ref{sec:ls}). 
Finally we report the correlation between the muon flux and atmospheric temperature (section~\ref{sec:corr}) before summarizing our results (section~\ref{sec:conclusions}).
 
Preliminary results of this analysis have been presented in \cite{dan11}.
 
\section{The Borexino Detector} 
\label{sec:bx}

\begin{figure}[t]
\centering
\includegraphics[width=.8\textwidth]{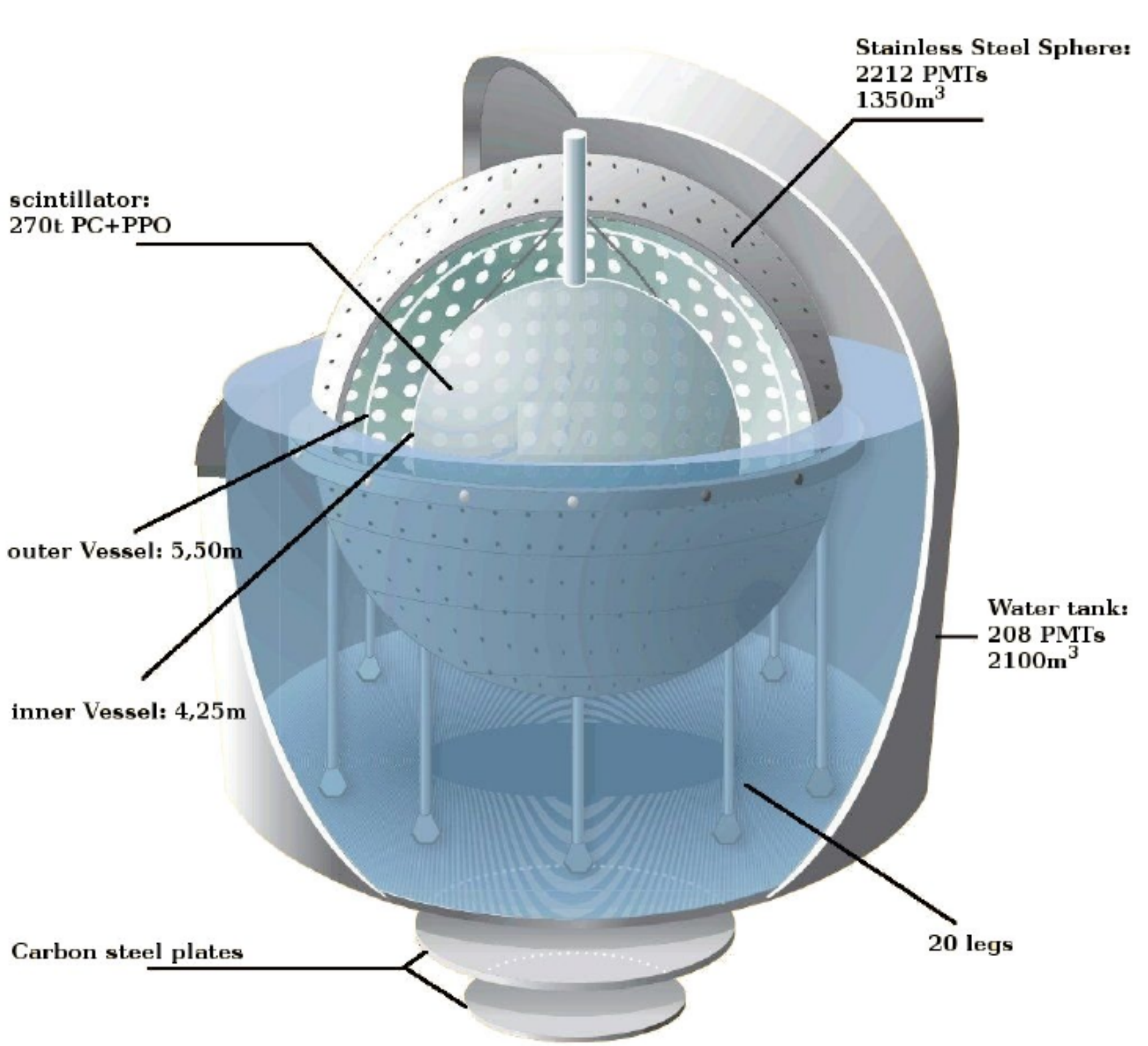}
\caption{Sketch of the Borexino detector.}
\label{fig:bx_sketch}
\end{figure}

The Borexino detector \cite{bx08det} is schematically shown in Figure~\ref{fig:bx_sketch}.
The active target for the analyses reported in this article is the Inner Detector (ID).
Its central scintillation volume consists of 278\,t
of ultra-pure PC (1,2,4-trimethylbenzene) doped with 1.5\,g/l of the fluor PPO (2,5-diphenyloxazole), contained in a spherical transparent 8.5\,m diameter nylon Inner Vessel (IV). 
It is shielded by two buffer layers consisting of PC with the light quencher DMP (dimethylphthalate). 
The surrounding 13.7\,m diameter Stainless Steel Sphere (SSS) 
holds 2212 inward-facing 8'' photomultiplier tubes (PMTs) 
that detect scintillation light from the central region. 
The ID is surrounded by a powerful muon Outer Detector (OD) \cite{bx11mv},
composed of a high domed steel tank of 18\,m diameter and 16.9\,m height filled
with 2\,100\,t of ultra-pure water and instrumented with 208 PMTs that detect the Cherenkov emission from cosmic ray muons.

\section{The Cosmic Muon Flux} 
\label{sec:flux}

The Gran Sasso underground laboratory consists of three experimental Halls, named A, B and C.
These are about 80\,m distant from each other and, in principle, could feature slightly different rock coverage.
Borexino is located in Hall C, while the cosmic muon flux has been measured previously by LVD in Hall A \cite{sel09} and by MACRO in Hall B \cite{macro95}. 
These experiments reported a flux of $(3.31\pm0.03)\cdot 10^{-4}$m$^{-2}$s$^{-1}$ and $(3.22\pm0.08)\cdot10^{-4}$m$^{-2}$s$^{-1}$, respectively. 
In both cases the acceptance is strongly dependent on the muon incidence angle, contributing significantly to the systematic error of the final result. 
When comparing available results it should also be noted that the flux measured in different years can reflect differences in the mean air temperature (see table~\ref{tab:cmp}).

This analysis is based on the first four years of Borexino data, taken between 16$^{\rm th}$ of May 2007 and 15$^{\rm th}$ of May 2011.
The CNGS (CERN Neutrino to Gran Sasso \cite{web-proj-cngs, Acquistapace}) neutrino beam introduces muon events that are a background for this analysis. 
Consequently all events in coincidence with the beam spills are discarded (see \cite{bx11mv} for details). 
The effective data set shows no prolonged or unevenly distributed downtime and accounts for a live time of 1063 days.
Muon detection in Borexino is performed with both the ID and OD, however here we disregard events that generated a signal in the OD only.
In this analysis the relevant cross-section for cosmic muons is therefore given by that of the SSS (146\,m$^2$), independent of the muon incoming angle. 
The corresponding total exposure of the data set is $\sim$1.55$\cdot10^{5}$ m$^2\cdot$d 
and includes a sample of $\sim$4.6$\cdot10^{6}$ muons.

The muons passing through the inner detector can be identified via three different methods. 
The first two are based on the detection of the Cherenkov light produced in the water: the light triggers the OD sub-system (MTF) or a cluster of hits is recognized within the time distribution of OD PMT hits (MCF). 
The third method (IDF) relies on the pulse shape identification of muon tracks among the point-like scintillation events detected by the ID. 
The detection efficiencies are 0.9925(2), 0.9928(2) and 0.9890(1) respectively.
Details on the muon tagging methods and on how the efficiencies have been evaluated can be found in \cite{bx11mv}.

We have measured the muon rate through the ID using all strategies at our disposal and achieved consistent results. 
The average muon rate is ($4310\pm2_{\textrm{stat}}\pm10_{\textrm{syst}}$) d$^{-1}$ where the systematic error reflects the uncertainty in the efficiency and possible threshold effects.
The rate corresponds to a cosmic muon flux of $(3.41\pm0.01)\cdot10^{-4}$m$^{-2}$s$^{-1}$, taking into account also the uncertainty in the SSS radius. 
This is the first measurement performed in Hall C and the first obtained with a spherical detector at LNGS.

\section{The Flux Modulation} 
\label{sec:modulation}

Air temperature increases during summer which lowers the average gas density.
The less dense medium allows a longer mean free path of the mesons and increases the fraction of them that decay to produce muons before their first interaction. 
As only these muons are energetic enough to traverse the rock coverage of an underground site, a correlation between the muon flux observed underground and the air temperature is expected.
We demonstrate such a correlation for the case of Borexino in section \ref{sec:corr}.
Temperature fluctuations can have maxima and minima that occur at different dates in successive years and short term effects that are expected to perturb the ideal seasonal variation. 
Therefore a simple sinusoidal behavior is to be considered only a first order approximation.

\begin{figure*}[t]
\centering
\includegraphics[width=\textwidth]{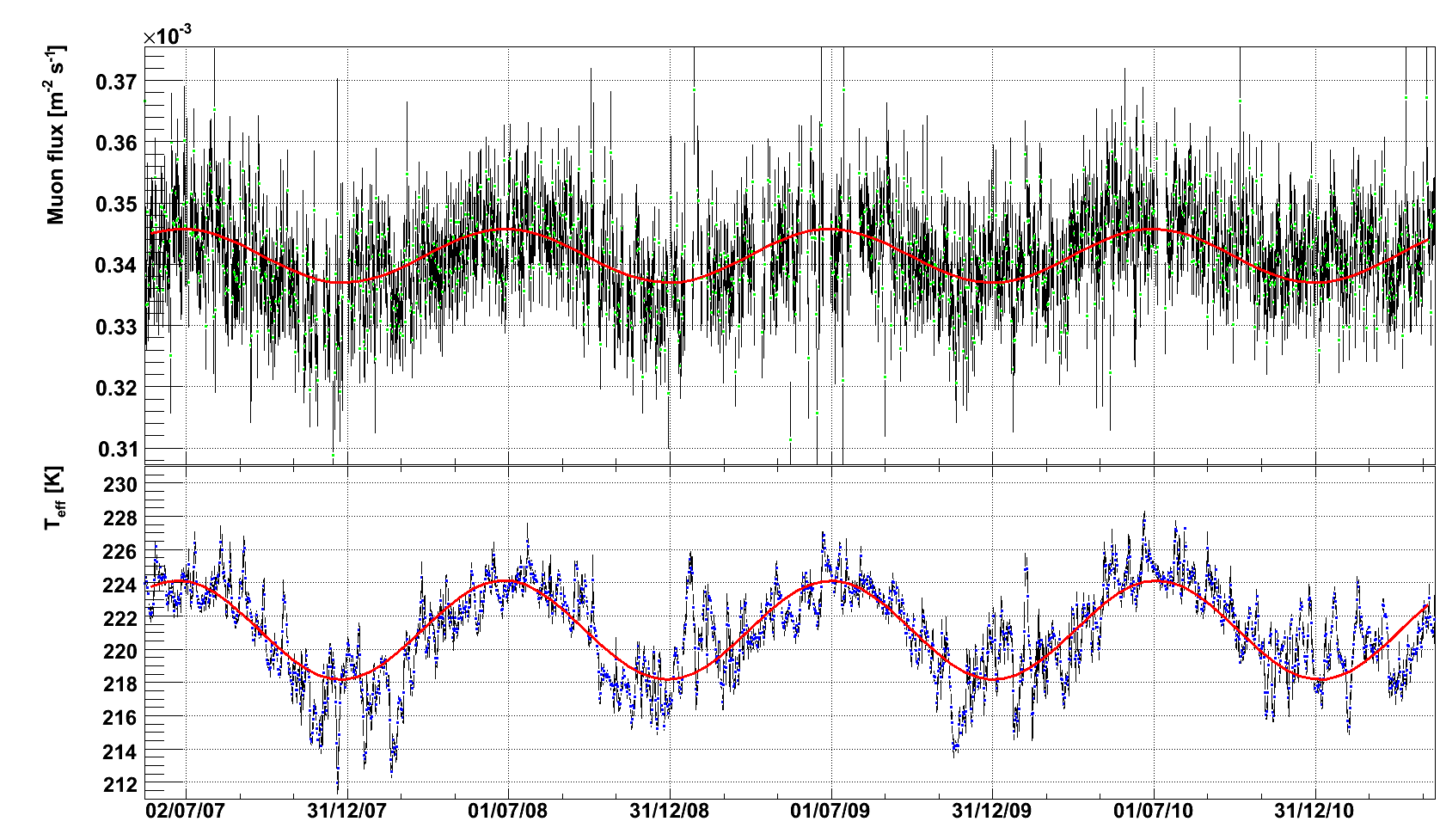}
\caption{Upper panel: cosmic muon signal measured by Borexino as a function of time. Lower panel: effective temperature, $T_{\textrm{eff}}$, computed using eq.~\ref{eq:teff} and averaging over the four daily measurements. Daily binning is used in both panels. The curves show the sinusoidal fit to the data (see text).}
\label{fig:temp_flux}
\end{figure*}

The muon flux measured day-by-day in Borexino is shown in figure~\ref{fig:temp_flux} (upper panel) for the 1329 days for which valid data were available.
A modulation is clearly visible. Fitting the distribution with the following function:
\begin{equation}
I_{\mu} = I_{\mu}^{0} + \Delta I_{\mu} =  I_{\mu}^{0} + \delta I_{\mu} \cos \left(\frac{2\pi}{T}(t-t_{0})\right)
\label{eq:seasonal_fit}
\end{equation}
we obtain an average intensity $I_{\mu}^{0}=(3.414\pm0.002_{\textrm{stat}})\cdot 10^{-4}$m$^{-2}$s$^{-1}$, consistent with the flux reported in section~\ref{sec:flux}, 
a period $T=(366\pm3)$ days, 
a modulation amplitude $\delta I_{\mu}=(4.4\pm0.2) \cdot 10^{-6}$m$^{-2}$s$^{-1}$, corresponding to $(1.29\pm0.07)\%$ 
and a phase $t_{0}=(179\pm6)$ days, corresponding to a maximum on the 28$^{\rm th}$ of June;
the Neyman's $\chi^{2}$/NDF is 1558/1325.

\begin{figure}[t]
\centering
\includegraphics[width=.65\textwidth]{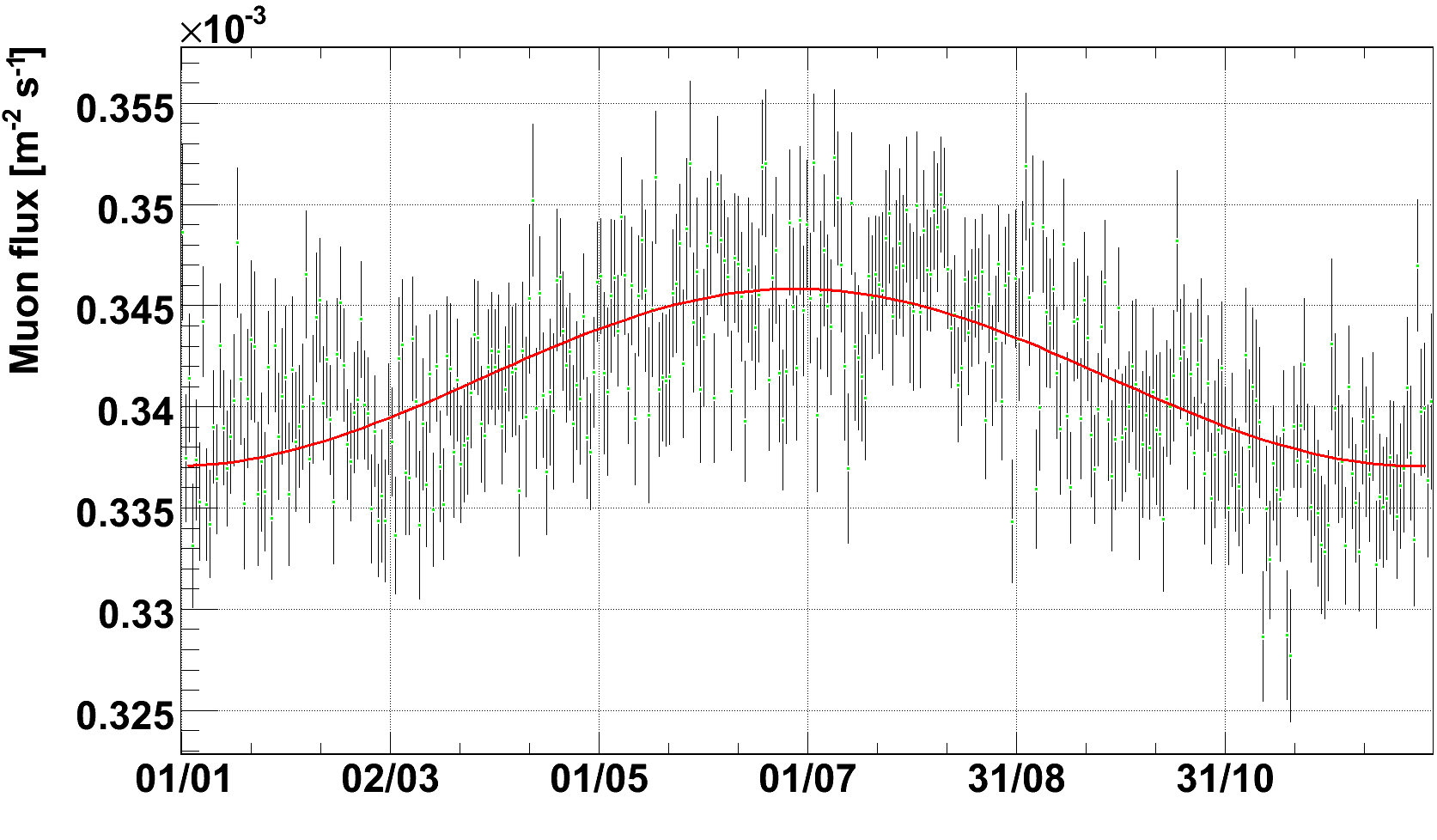}
\caption{Cosmic muon flux: four years data set folded onto a one year period. Daily binning. The curve shows the
sinusoidal fit to the data (see text).}
\label{fig:seasonal_1y}
\end{figure}

An alternative approach is to project and average the four years data set into one single year, as shown in figure~\ref{fig:seasonal_1y}.
Fitting again with eq.~\ref{eq:seasonal_fit}  but with the period fixed to one year, we obtain consistent rate and amplitude.
The phase is $t_{0}= (179\pm3)$ days. The $\chi^{2}$/NDF of the fit is 442/362.

\section{The Atmospheric Model} 
\label{sec:model}

Deviations from the average muon flux that is measured underground, $\Delta I_{\mu} (t)  = I_{\mu} (t) -I^0_{\mu}$, can be related to variations from the average atmospheric temperature at a given altitude $X$, $\Delta T(X, t) = T(X, t) - T^0(X) $ (from \cite{minos10}). 
Considering every altitude layer, the net effect can be written as:
\begin{equation}
\Delta I_{\mu} = \int_{0}^{\infty}dXW(X)\Delta T(X)
\label{eq:seasonal}
\end{equation}
where $W(X)$ (see appendix~\ref{app:weights}) reflects the altitude dependence of the production of mesons in the atmosphere and their decay into muons that can be observed underground. The integral extends over atmospheric depth from the altitude of muon production to the ground. 

The atmosphere can be described by many layers 
with a continuos distribution of temperature and pressure.
A possible parametrization (\cite{minos10} and with more details \cite{gra10}) considers the atmosphere as an isothermal body with an effective temperature, $T_{\textrm{eff}}$, obtained from a weighted average over atmospheric depth:
\begin{equation}
T_{\textrm{eff}} = \frac{\int_0^{\infty}dXT(X)W(X)}{\int_0^{\infty}dXW(X)}\simeq\frac{\sum^N_{n=0} \Delta X_nT(X_n) W (X_n)}{\sum^N_{n=0} \Delta X_n W (X_n)}
\label{eq:teff}
\end{equation}
where the approximation may be done considering that the temperature is measured at discrete atmospheric levels, $X_n$.

\begin{figure}[t]
\centering
\includegraphics[width=\columnwidth]{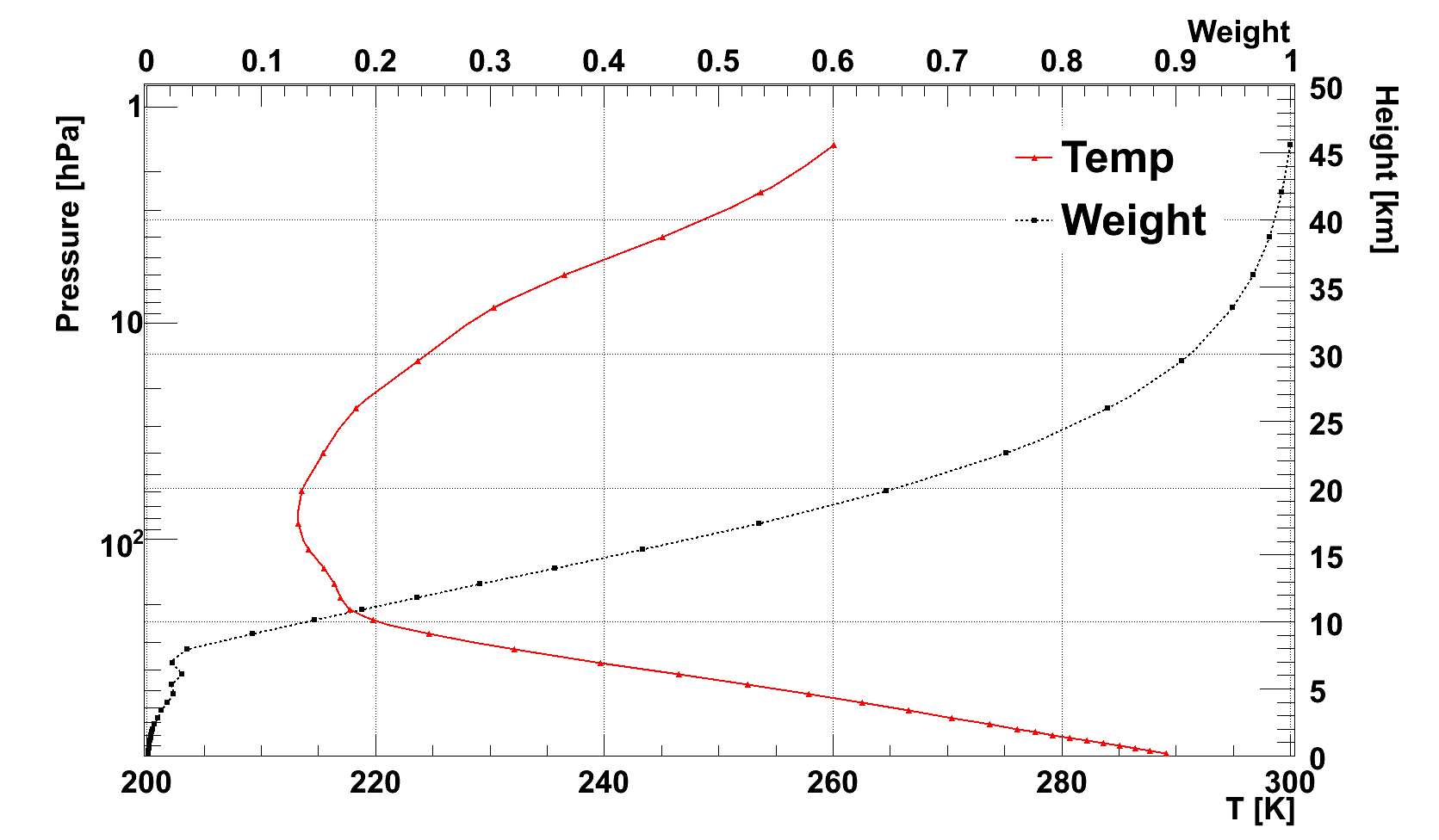}
\caption{Average temperature (solid red line) \cite{ecmwf} and normalized weight $W(X)$ (black dashed line) as a function of pressure levels computed at the LNGS site. The right vertical axis shows the altitude corresponding to the pressure on the left vertical axis.}
\label{fig:tavg_weight}
\end{figure}
 
Figure~\ref{fig:tavg_weight} shows the temperature in the atmosphere for the LNGS site and the weight function, $W$, as functions of the pressure levels.
As can be seen, the higher layers of the atmosphere are given higher weights, as it is in these layers that most of the muons energetic enough to reach underground sites are produced.
Muons produced at a lower altitude will be on average less energetic and a larger fraction of them will lie below threshold ($E_{\textrm{thr}}$). 

We may also define the  ``effective temperature coefficient'', $\alpha_T$, which quantifies the correlation effect that is discussed in section~\ref{sec:corr}:
\begin{equation}
\alpha_T = \frac{T^0_{\textrm{eff}}}{I^0_{\mu}} \int_0^{\infty}dXW(X)
\label{eq:alphat}
\end{equation}
such that Eq.~\ref{eq:seasonal} may be written:
\begin{equation}
\frac{\Delta I_{\mu}}{I^0_{\mu}} = \alpha_T \frac{\Delta T_{\textrm{eff}}}{T^0_{\textrm{eff}}}
\label{eq:var_at}
\end{equation}

\section{Temperature Modulation} 
\label{sec:temp}

The temperature data was obtained from the European Center for Medium-range Weather Forecasts (ECMWF)\cite{ecmwf}
which exploits different types of observations (e.g. surface, satellite, and upper air sounding) at many locations around the planet, 
and uses a global atmospheric model to interpolate to a particular location. 
In our case, the precise coordinates of the LNGS underground halls have been used: 13.5333$^\circ$\,E, 42.4275$^\circ$\,N.
Atmospheric temperature is provided by the model at 37~discrete pressure levels in the [1-1000]\,hPa range (1 hPa = 1.019 g/cm$^2$), 
four times a day at 00.00 h, 06.00 h, 12.00 h, and 18.00 h
\footnote{The analysis in \cite{macro97} and \cite{sel09} used data from the air soundings performed by the Aeronautica Militare Italiana (AM) \cite{am} near the military base of Pratica di Mare (12.44$^\circ$\,E, 41.65$^\circ$\,N), about 130\,km away from the lab. 
Aside to referring to a somewhat different location, that data set --- probably the best available at the time of \cite{macro97} --- is significantly incomplete if compared to the one from ECMWF, both for number of measurements and for atmospheric depth coverage. 
We therefore used this data set only as a cross-check of the analysis based on the ECMWF data set, yielding consistent results.}.
Based on this data set, $T_{\textrm{eff}}$ was calculated using eq.~\ref{eq:teff} four times a day. 
The four results were then averaged, and the variance of the four values was used to estimate the uncertainty in the mean.

Figure~\ref{fig:temp_flux} (lower panel) shows the daily values of $T_{\textrm{eff}}$ for the four year period considered. 
A simple average gives $T^0_{\textrm{eff}}  = 220.99$\,K, 
while the fit with a function analogous to eq.~\ref{eq:seasonal_fit} returns $T^0_{\textrm{eff}}=(221.153\pm0.007)$\,K, 
amplitude $(2.98\pm0.01)$\,K corresponding to 1.35\,\%, period $(369.2\pm0.2)$ days and phase $(174.0\pm0.4)$ days. 
These are very similar to the corresponding fit results of the muon flux data set, discussed in section~\ref{sec:modulation}.
The $\chi^2$/NDF = 46010/1457 confirms that the sinusoidal behavior is only a first order approximation.
Aside from small scale fluctuations, additional winter maxima can be observed which can be ascribed to the known meteorological phenomenon of the Sudden Stratospheric Warmings (SSW \cite{osp09}) and whose effect is sometimes comparable in amplitude with the underlying seasonal modulation.
In order to disentangle the seasonal dependence from sub-leading effects, the method of Lomb-Scargle has been used.

\section{Lomb-Scargle analysis}
\label{sec:ls}

Lomb-Scargle (LS) periodograms \cite{lom76,sca82} are a common method to analyze a binned data set for periodical modulations of the type
\begin{equation} \label{EqBasMod}
N(t) = N_0\cdot\left(1+A\cdot\sin(2\pi t/T+\varphi)\right).
\end{equation}
Here, $N(t)$ is the expected event rate at time $t$, while $N_0$ represents the mean rate, $A$ indicates the relative amplitude of the modulation, $T$ describes its period and $\varphi$ the phase relative to the start of the measurement. The power, $P$, for a particular modulation period, $T$, is defined as the weighted difference between the number of events, $N(t_i)$, in every data bin, $i$, and the weighted  mean value, $N_0$, with cosine and sine functions that oscillate with an investigated period, $T$:
\begin{eqnarray}
P = \frac{1}{2\sigma^2}\left( \frac{\left[\sum_{i=1}^n w_i(N(t_i)-N_0)\cos\tau_i\right]^2 }{\sum_{i=1}^n w_i\cos^2\tau_i} + \frac{\left[\sum_{i=1}^n w_i(N(t_i)-N_0)\sin\tau_i\right]^2 }{\sum_{i=1}^n w_i\sin^2\tau_i}\right),
\label{eq::lombscargle}
\end{eqnarray}
where $n$ is the number of bins and $t_i$ the time at which the data corresponding to bin $i$ was acquired. The weights $w_i=\sigma_i^{-2}/\langle \sigma_i^{-2} \rangle$ are the inverse squares of the statistical uncertainties of individual bins, $N(t_i)$, divided by their average value. Accordingly, $\sigma^2$ is the weighted variance of the data. The phase of the sine and cosine weights, $\tau_i$, is defined as \cite{sno05}
\begin{eqnarray}
\tau_i = 2\pi\cdot\frac{t_i-t_p}{T} \qquad {\rm with} \qquad t_p = \frac{T}{4\pi}\arctan\left(\frac{\sum_{i=1}^N w_i\sin\left(4\pi \frac{t_i}{T}\right)}{\sum_{i=1}^N w_i\cos\left(4\pi \frac{t_i}{T}\right)}\right).
\end{eqnarray} 
As the quadratic sums of both cosine and sine are used in eq.\,(\ref{eq::lombscargle}), the result is independent of the modulation phase as long as the modulation period is shorter than the overall measurement time. 

\begin{figure*}
\begin{minipage}{0.48\textwidth}
\includegraphics[width=\textwidth]{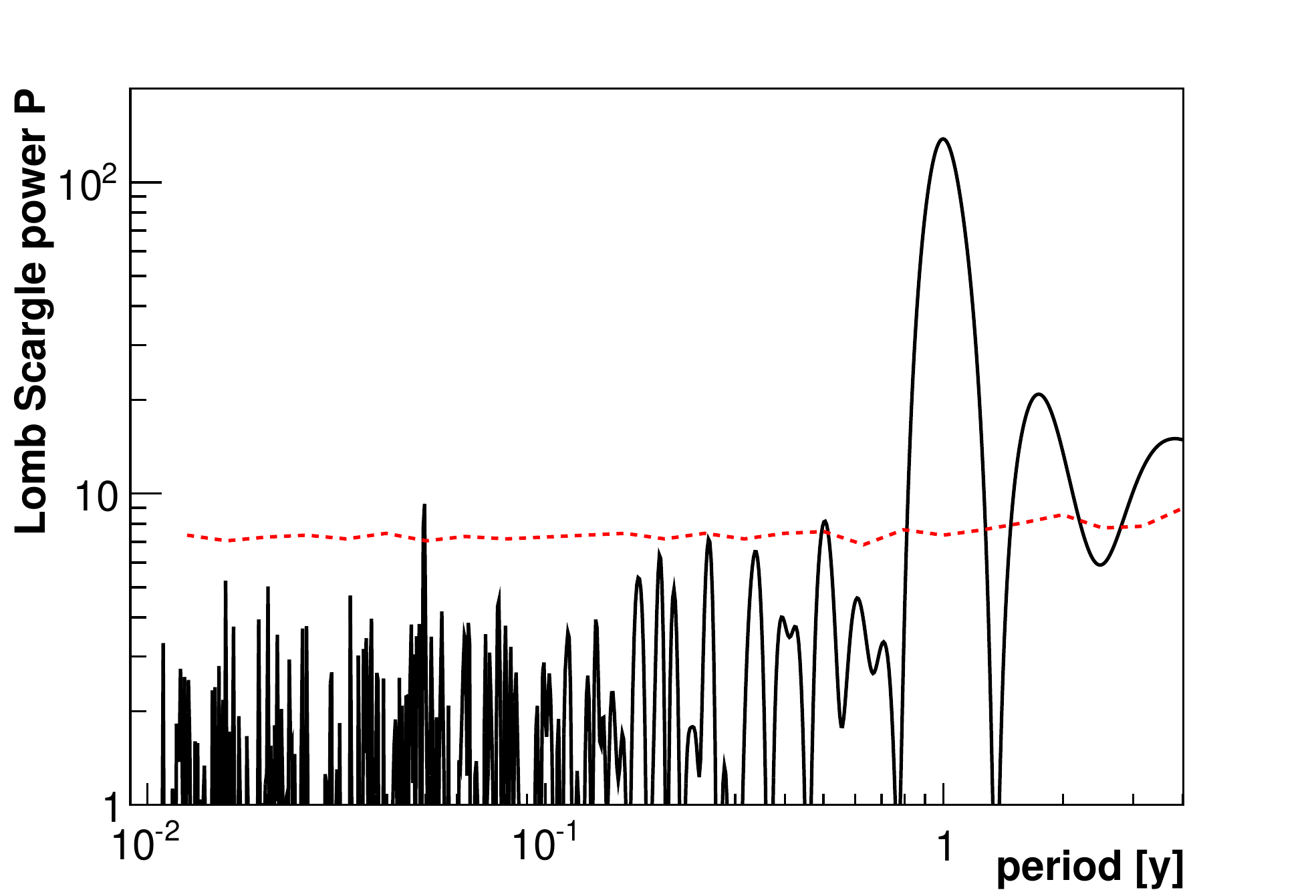}
\caption{LS periodogram of the muon data. The dashed line indicates the detection threshold (3$\sigma$).}
\label{fig::muon_ls}
\end{minipage}
\hfill
\begin{minipage}{0.48\textwidth}
\includegraphics[width=\textwidth]{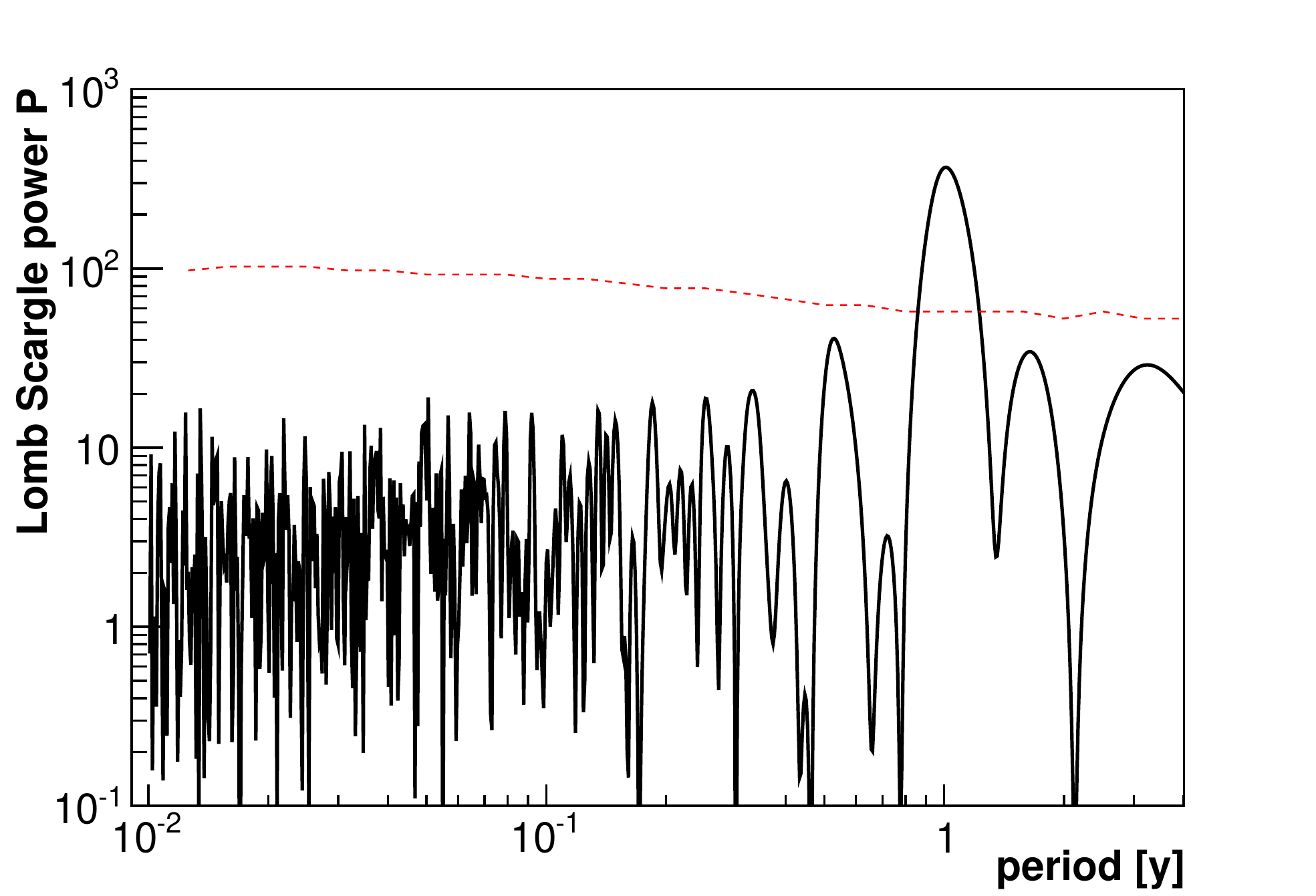}
\caption{LS periodogram of the temperature data. The dashed line indicates the detection threshold (3$\sigma$).}
\label{fig::temp_ls}
\end{minipage}
\end{figure*}

Figure~\ref{fig::muon_ls} shows an LS periodogram of the four-year muon data acquired in Borexino. The LS power, $P$, of a given modulation is primarily a function of its amplitude, $A$. Statistical fluctuations of the bin content will alter both the maximum, $P$, generated by white noise and the exact value observed for the actual modulation. To assess the significance of a modulation discovery, it is therefore necessary to know the statistical fluctuations of both the white noise level and the signal height.

Using the known detector live time distribution and muon rate, we have performed realistic Monte Carlo (MC) simulations of muon sample time distributions. In a first step, $10^4$ MC data samples with a constant muon rate have been simulated. From the corresponding $P$ spectra we can estimate the probability that an apparent periodic modulation appears as a result of white noise. For a given period $T$, the $3\sigma$ detection threshold $P_\mathrm{thr}(T)$ is set so that 99.7\,\% of all white noise samples feature lower values of $P$.
This threshold is indicated by the dashed line in figure~\ref{fig::muon_ls}. Over a broad range of periods, we obtain values of $P_\mathrm{thr} \approx 7$ ($3\sigma$).

Based on a further set of MC data samples that contain periodic modulations, given values of $P$ can be matched to modulation amplitudes and vice versa: $P_\mathrm{thr}$ corresponds to a relative amplitude of $\sim$0.3\,\%, the exact value depending on statistical fluctuations. Compared to threshold, the $P$ peak corresponding to the annual modulation is highly significant. The maximum of $P=140$ is reached for a period of 364 days. From MC simulation, one can associate this LS power to an amplitude of $(1.20\pm0.05)\,\%$. 

In figure~\ref{fig::muon_ls}, several secondary peaks are visible: two peaks at $T=0.05$ years and $T=0.5$ years are just above detection threshold, while the largest secondary peak is at $T=1.7$ years featuring $P\approx20$, corresponding to an amplitude of about 0.4\,\%. The appearance of these additional peaks indicates that there are deviations of the modulation pattern from a simple year-long sinusoidal. Dividing the data sample by the main modulation allows to investigate whether the side-peaks are reflections of the annual modulations at harmonic frequencies. This procedure increases the significance of the peak at $T=0.5$ years to $P=15$, while both the peaks at $T=0.05$ years and $T=1.7$ years are pushed below detection threshold. To further investigate this semi-annual modulation, a second sinusoid was added to the direct fit to the data set (eq.~\ref{eq:seasonal_fit}). The period of the second sinusoid fits to $T=(179\pm2)$ days, and the amplitude to $A=(0.37\pm0.07)$\,\%. The fit result for the main (1-year) oscillation remains unaffected within uncertainties.
We conclude that the modulation found in the data is best described by the superposition of two sinusoidal terms with the semi-annual component slightly distorting the rising and
falling flanks of the main annual component.

The LS analysis can also be applied to the temperature data. The resulting periodogram of figure~\ref{fig::temp_ls} features again a prominent peak at $T=368$\,days. The observed value of $P=367$ is compatible with the modulation amplitude of 1.35\,\% (section~\ref{sec:temp}) with 90\,\% confidence level. 
The observation threshold (3$\sigma$) for the temperature data set is $P_\mathrm{thr}(1\,\rm y)=58$; none of the secondary peaks is significant at 3$\sigma$ level. The peak corresponding to the semi-annual sub-modulation identified for the muons has a $\sim$2.2$\sigma$ significance in the temperature data.

\section{Correlation} 
\label{sec:corr}

\begin{figure}[t]
\centering
\includegraphics[width=\columnwidth]{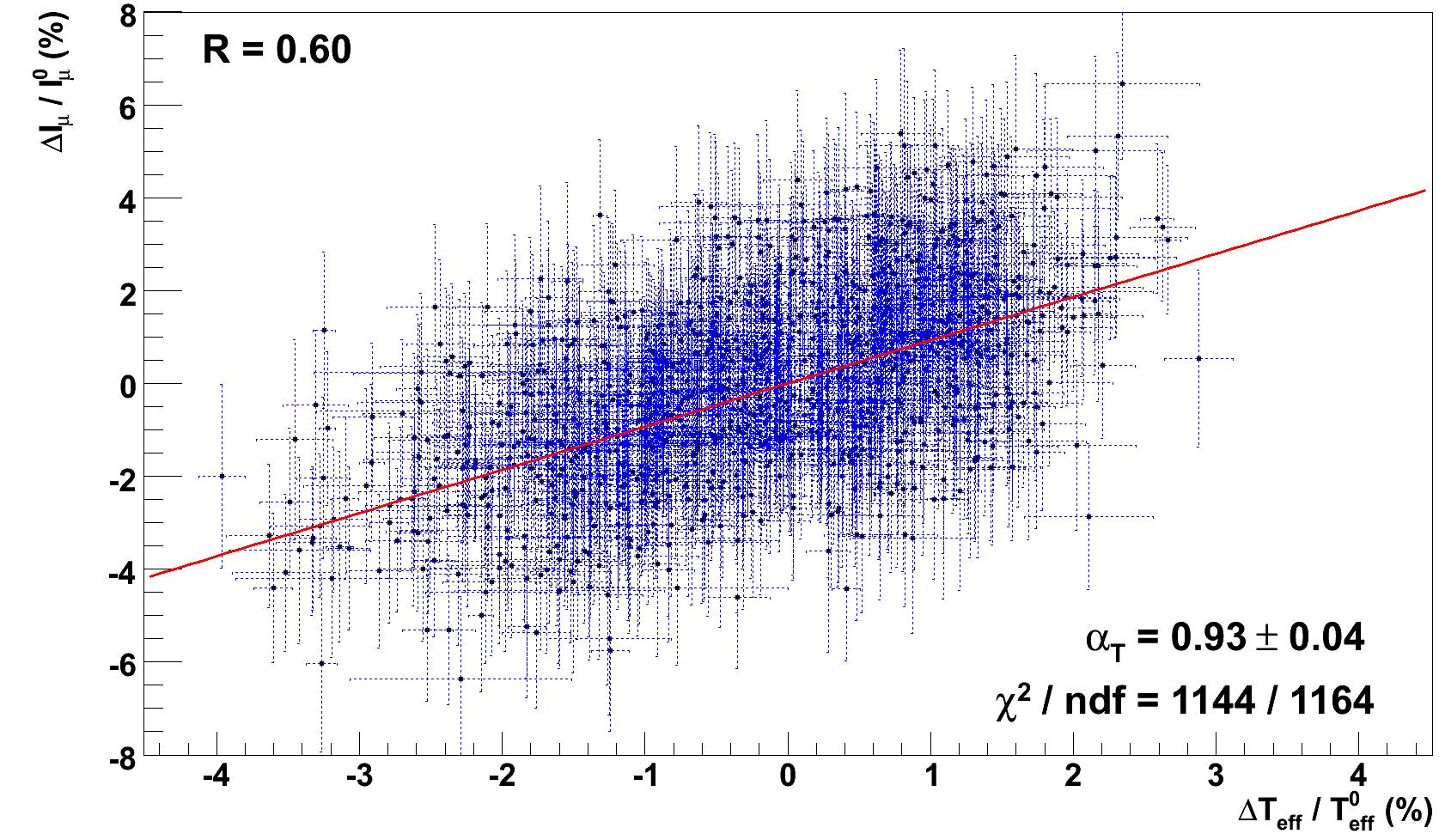}
\caption{$\Delta I_{\mu}/I^0_{\mu}$ vs. $\Delta T_{\textrm{eff}}/ T^0_{\textrm{eff}}$. 
Each point corresponds to one day of data.}
\label{fig:correlation}
\end{figure}

Figure~\ref{fig:temp_flux} shows the correlation between fluctuations in the atmospheric temperature and the cosmic muon flux. 
To quantify such a correlation we plot $\Delta I_{\mu}/I^0_{\mu}$ vs $\Delta T_{\textrm{eff}}/T^0_{\textrm{eff}}$ for each day in figure~\ref{fig:correlation}. 
Only days with duty cycle $\ge 50\%$ are included for a total of 1165 days.
The correlation coefficient (R-value) between these two distributions is 0.60 indicating a positive correlation.
We now want to determine the effective temperature coefficient (eq.~\ref{eq:alphat}).
We perform a linear regression accounting for error bars on both axes using a numerical minimization method. 
As a result we obtain $\alpha_T = 0.93 \pm 0.04_{\textrm{stat}}$ with $\chi^2$/NDF = 1144/1164.
This result is consistent and features smaller errors when compared to $\alpha_T = 0.91 \pm 0.07$, the previous measurement by MACRO at Gran Sasso \cite{macro03}. 

We have performed the following tests to check for systematic uncertainties:
\begin{itemize}
\item
$I^0_{\mu}$ and $T^0_{\textrm{eff}}$ have been computed in different ways: averaging $I_{\mu}$ and $T_{\textrm{eff}}$ values over the available data set; from the fit to the four year data set with the sinusoidal functions as in eq.~\ref{eq:seasonal_fit} and figure~\ref{fig:temp_flux}; from a fit of the same data set with a constant function. In addition $T^0_{\textrm{eff}}$ has been computed including or excluding the days for which no corresponding muon flux was available. 
\item The analysis has been performed on a moving two-year sub-period checking the stability of the result. 
\item We have varied the requirement of including only days with duty cycle $\ge50\%$ in the range $[\ge20\%, \ge80\%]$. 
\item We have considered $E_{\textrm{thr}}=1.833$ TeV from \cite{gra10} and $E_{\textrm{thr}}=1.3$ TeV as in \cite{macro97,sel09} for the computation of $T_{\textrm{eff}}$ (see also appendix~\ref{app:weights}).
\item We ran the analysis substituting the air temperature data set from ECMWF with that from Aer. Mil. Italiana (see footnote in section~\ref{sec:temp}) used in \cite{macro97,sel09}.
\item We recomputed $T_{\textrm{eff}}$ using weights that account for muon production only from pion decay, i.e. neglecting the kaon contribution as done in \cite{macro97,sel09} (see appendix~\ref{app:weights}).
\end{itemize}  
In all cases we found our result to be stable, so we believe that the systematic uncertainty is small compared to the statistical error from the fit.

\begin{figure}[t]
\centering
\includegraphics[width=\columnwidth]{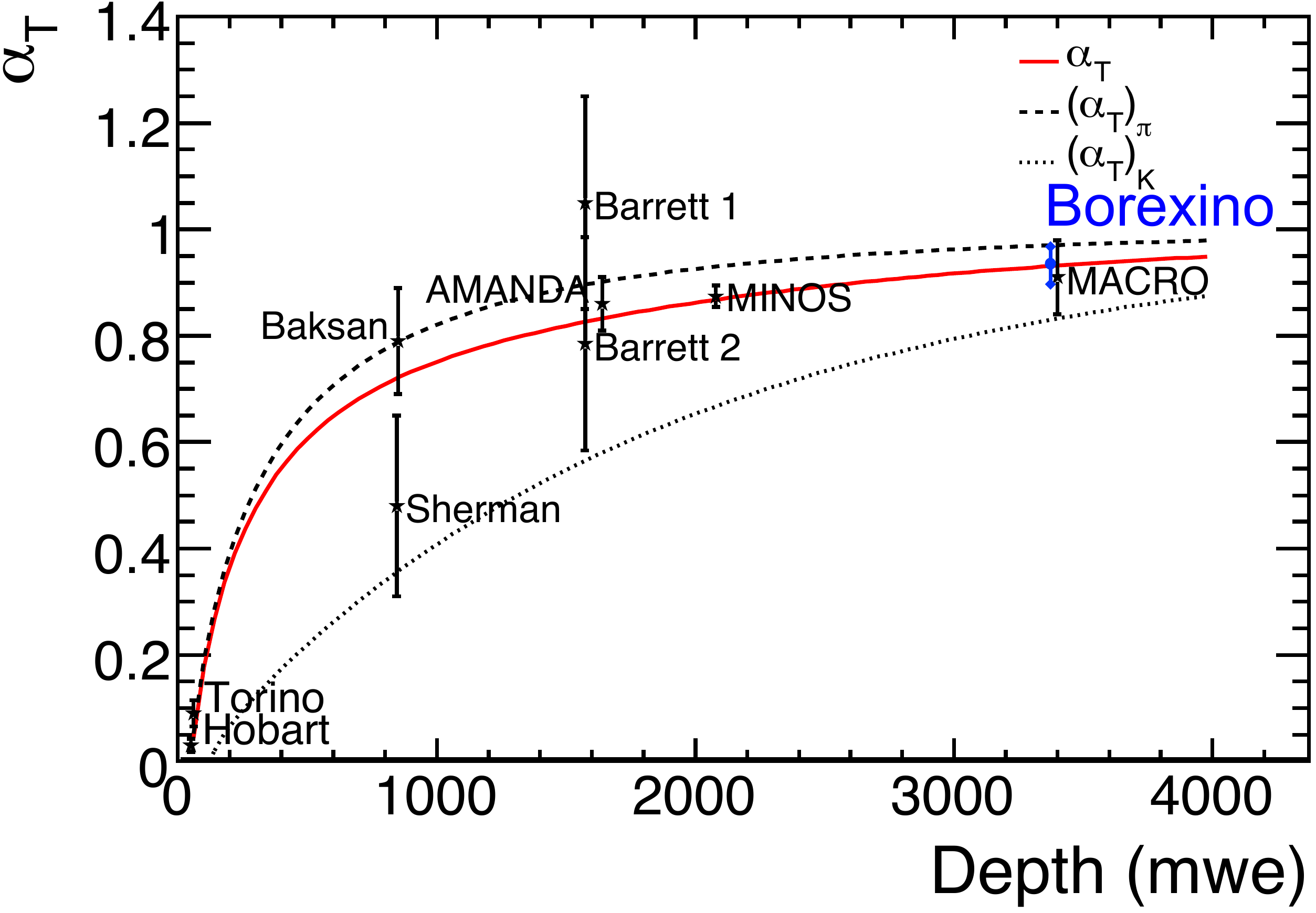}
\caption{Measured values for the effective temperature coefficient, $\alpha_{T}$, at varying site depths. 
The results from this analysis (in blue) as well as those from different experiments are presented. 
The red line is the value predicted including muon production by pions and kaons. 
The dashed lines account for one production mechanism only. See \cite{minos10} and refs therein for details. }
\label{fig:alpha_t}
\end{figure}
   
Figure~\ref{fig:alpha_t} shows the measured values for $\alpha_{T}$, with the details summarized in table~\ref{tab:cmp}. 
This value has been predicted for different site depths following \cite{minos10}. 
As shown in figure~\ref{fig:alpha_t},  $\alpha_{T}$ asymptotically approaches unity with increasing site depth. This is because the air-density-independent contribution to the muon signal originating from mesons which have interacted before decaying is progressively left below threshold. 
At LNGS $\alpha_T$ is expected to be $0.92\pm0.02$ (considering muon production from both pions and kaons) in good agreement with the result from this analysis.
The systematic uncertainty in this prediction was found by modifying the input parameters for the computation according to their uncertainties and recalculating.

\begin{table}
\begin{tabular}{|l|cc|ccc|}
\hline
Site [km w.e.]&S. Pole(1.6)&Soudan(2.1) &{\small LNGS-A}(3.8) &{\small LNGS-B}(3.8) &{\small LNGS-C}(3.8)\\
Experiment&{\small ICECUBE\cite{icecube}}&{\small MINOS\cite{minos10}}&{\small LVD\cite{sel09}}&{\small MACRO\cite{macro03}}&Borexino\\
$E_{\textrm{thr}}$[GeV]&466&730&1833&1833&1833\\
Rate [$10^3\mu$/d]&100000&40&8&9&4\\
Meas. time [y]&2007-11[4y]&2003-08[5y]&2001-08[8y]&1991-97[7y]&2007-11[4y]\\
Accept. [$\cos(\theta)$] &any&[0.05-0.92]&$>$0.5&$>$0.3&any\\
\hline
{\small $I_{\mu}[10^{-4}m^{-2}s^{-1}]$}&&&$3.31\pm0.03$&$3.22\pm0.08$&$3.41\pm0.01$\\
Modul. Ampl.&&&1.5\%&&1.3\%\\
Period (days)&&&$367\pm15$&&$366\pm3$\\
Phase (days) &&&$185\pm15$&&$179\pm6$\\
&&&(Jul $4^{\rm th}$)&&(Jun $28^{\rm th}$)\\
\hline
Binning&daily&daily&daily&monthly&daily\\
Air Data&NASA-AIRS&ECMWF&Aer.Mil.&Aer.Mil.&ECMWF\\
$T_{\textrm{eff}}$ model&$\pi$+K&$\pi$+K&$\pi$-only&$\pi$-only&$\pi$+K\\
Correlation&n.p.&0.90&0.53&0.91&0.62\\
$\alpha_T$(meas.)&$0.860\pm0.010$&$0.873\pm0.009$&$--$&$0.91\pm0.07$&$0.93\pm0.04$\\
$\alpha_T$(pred.)&$\sim0.83$&$0.864\pm0.024$&$0.92\pm0.02$&$0.92\pm0.02$&$0.92\pm0.02$\\
\hline
\end{tabular}
\caption{Comparison of the different analyses of the muon seasonal modulation and correlation with temperature by some existing underground experiments.}
\label{tab:cmp}
\end{table}

With a longer exposure we expect to measure $\alpha_T$ with better precision, opening the way to an indirect determination of the K/$\pi$ ratio in the interaction of primary cosmic rays in the atmosphere with the method detailed in \cite{minos10, gra10} and probing a complementary energy region compared with existing accelerator experiments.

\section{Conclusions}
\label{sec:conclusions}

Borexino has reached four years of continuous data taking at LNGS under a rock coverage of 3800\,m\,w.e.
Due to the spherical geometry of the detector, we have measured the underground cosmic muon flux with reduced systematics:
$(3.41\pm0.01)\cdot10^{-4}$m$^{-2}$s$^{-1}$.
We also have observed a seasonal modulation of the flux and measured the amplitude to be ($1.29\pm0.07$)\,\% and the phase to be ($179\pm6$) days corresponding to a maximum on the $28^{\rm th}$ of June.
To invesitgate the correlation between air temperature variations and changes in the muon flux, 
we have obtained air temperature data from global atmospheric models for the same four years period. 
We showed that the annual modulation is also present in the effective temperature data, with oscillation parameters compatible with those of the muon modulation.
We then showed that the two data sets are positively correlated ($R=0.60$) and we measured the effective temperature coefficient to be $\alpha_T = 0.93 \pm 0.04$. 
This result is compatible with theoretical expectations and is an improvement in precision from previous measurements at Gran Sasso.

\acknowledgments

We thank the funding agencies: INFN (Italy), NSF (USA), BMBF, DFG and MPG (Germany), NRC (Russia), MNiSW (Poland), and we acknowledge the generous support of the Gran Sasso National Laboratory.
We thank E.W. Grashorn of CCAPP, Ohio State University for insightful discussions and S.M. Osprey of NCAS, University of Oxford (UK) for promptly providing ECMWF air temperature data interpolated on the LNGS coordinates.

\appendix

\section{Effective temperature weight function}
\label{app:weights}

The weight $W(X)$ used in eq.~\ref{eq:teff} can be written as the sum $W_{\pi}+W_{K}$, representing the contribution of pions and kaons to the overall variation in muon intensity:
\begin{equation}
W^{\pi,K}(X) \simeq \frac{(1-X/\Lambda'_{\pi,K})^2 e^{-X/\Lambda_{\pi,K}} A^1_{\pi,K}}{\gamma+(\gamma+1)B^1_{\pi,K}K(X)(\langle E_{\textrm{thr}}\cos\theta\rangle /\epsilon_{\pi,K})^2}
\label{eq:weights}
\end{equation}
where
\begin{equation}
K(X) \equiv \frac{(1-X/\Lambda'_{\pi,K})^2}{(1-e^{-X/\Lambda'_{\pi,K}})\Lambda'_{\pi,K}/X}
\label{eq:KX}
\end{equation}
The parameters $A^1_{\pi,K}$ include the amount of inclusive meson production in the forward fragmentation region, the masses of mesons and muons, and the muon spectral index $\gamma$; 
the input values are $A^1_{\pi} = 1$ and $A^1_K = 0.38 \cdot r_{K/\pi}$, where $r_{K/\pi}$ is the $K/\pi$ number ratio. 
The parameters $B^1_{\pi,K}$ reflect the relative atmospheric attenuation of mesons; 
the threshold energy, $E_{\textrm{thr}}$, is the energy required for a muon to survive to a particular underground depth and $\theta$ is the angle between the muon and the vertical directions; 
the attenuation lengths for the cosmic ray primaries, pions and kaons are $\Lambda_N$, $\Lambda_{\pi}$ and $\Lambda_K$ respectively with $1/\Lambda'_{\pi,K}= 1/\Lambda_N-1/\Lambda_{\pi,K}$. 
The meson critical energy, $\epsilon_{\pi,K}$, is the meson energy for which decay and interaction have an equal probability. 
The value of $\langle E_{\textrm{thr}} \cos \theta \rangle$ used here is the median of the distribution. 
The values for these parameters can be found in table 1 of \cite{minos10}, with the exception of $\langle E_{\textrm{thr}} \cos \theta \rangle$ which is site dependent and is found by MC simulations.
At LNGS $\langle E_{\textrm{thr}} \cos \theta \rangle$ = 1.833 TeV according to \cite{gra10}.
The dependence of $W(X)$ on $E_{\textrm{thr}}$ is however moderate.

\clearpage

\end{document}